\documentstyle[aps,prl,multicol,epsf]{revtex}

\def\bra#1{\langle #1 |}
\def\ket#1{| #1\rangle}

\def\Tr{{\rm Tr}~}

\def\R{\hbox{\rm I \kern-5pt R}}

\title{Entangled Mixed States and Local Purification}

\author{Adrian Kent}

\address{Department of Applied Mathematics and Theoretical Physics,
University of Cambridge,\\ 
Silver Street, Cambridge CB3 9EW, U.K.}

\date{28 May 1998; revised 23 August 1998}
\begin{document}
\maketitle
\begin{abstract}
Linden, Massar and Popescu have recently given an optimization
argument to show that a single two-qubit Werner state, or any other 
mixture of the maximally entangled Bell states, cannot be 
purified by local operations and classical communications. 
We generalise their result and give a simple explanation. 
In particular, we show that no purification scheme using local operations 
and classical communications can produce a pure singlet 
from any mixed state of two spin-$1/2$
particles.  More generally, no such scheme can produce a maximally
entangled state of any pair of finite-dimensional systems
from a generic mixed state.  
We also show that the Werner states belong to a large class of
states whose fidelity cannot be increased by such a scheme.  
\end{abstract}


\begin{multicols}{2}

\newcommand\mathC{\mkern1mu\raise2.2pt\hbox{$\scriptscriptstyle|$}
                {\mkern-7mu\rm C}}                     
\newcommand{\mathR}{{\rm I\! R}} 

The relationship between quantum entanglement and locality has been
a source of great theoretical interest ever since the discovery
of Bell's theorem, and new subtleties continue to be discovered.
Several practical applications of entanglement have also 
been proposed,\cite{BBCJPW,E,DEJMPS} 
most of which require that separated parties
share a fixed maximally entangled state --- conventionally, 
the Bell singlet state of two spin-$1/2$ particles.
Since no preparation method is perfect and no transmission channel is
noiseless, in practice we expect the parties to share entangled 
mixed states rather than pure singlets.  
This raises a problem which is also of independent theoretical 
interest: how can singlets be extracted from shared entangled 
mixed states?  

One solution, at least in principle, is to use the ``entanglement
purification'' scheme of Bennett et al.\cite{BBPSSW}.  
But as this scheme relies on carrying out
collective measurements on a large number of shared 
states, and produces perfectly pure singlets only when the number of 
shared states is infinite, it is natural to ask whether there is 
any simpler and 
more practical method.  In particular, it is natural to ask 
whether there is any way of purifying individual entangled mixed states
by a local purification scheme --- that is, a scheme 
which involves only local measurements and unitary operations on the 
two entangled particles, together with classical communications
between the parties.  

Linden, Massar and Popescu (LMP) have recently addressed this question,
showing that the answer is negative for the entangled 
Werner states\cite{W} in two dimensions --- i.e., for rotationally 
invariant mixed states of two spin $1/2$ particles whose fidelity to
the Bell singlet is greater than $1/2$.  More generally, they show
that the fidelity of a Werner state --- i.e. the proportion of 
singlets --- cannot be increased by local purification schemes, 
and that no mixture of the maximally entangled Bell states
of two spin-$1/2$ particles can be purified by such schemes
to a state whose entanglement of formation is greater.  
As LMP comment, these results may at first sight seem surprising,  
given that singlets {\it can} be produced from 
non-maximally entangled pure states by individual 
measurements.\cite{BBPS}  One might have 
conjectured that the same 
should be true of mixed states, at least if they are sufficiently
close to being singlets.  LMP's argument shows that this is not 
true, but does not give any clear intuition as to why pure and 
mixed states differ in this respect, and leaves open the possibility
that generic mixed states of spin-$1/2$ 
particles, or mixed states of higher-dimensional systems, 
might be locally purifiable.  

In this Letter we give a simple argument which generalises LMP's
results and explains why 
individual mixed states cannot generally be completely purified. 
We show that no local purification scheme can produce a pure singlet
from a single copy of any entangled mixed state of two spin $1/2$ 
particles.  More generally, we show that no local
purification scheme can produce a maximally entangled state 
from a generic mixed state entangling any two finite-dimensional 
systems.  Finally, we show that the Werner and Bell-diagonal states 
belong to a large class of states whose fidelity to a given state
cannot be increased by any local purification scheme.

Consider first a general mixed state of two spin-$1/2$ particles,
\begin{equation}
\label{m} 
\rho = \sum_{i=1}^{n} p_i \ket{\psi_i } \bra{\psi_i} \, , 
\end{equation} 
where the $\ket{\psi_i} \in C^2 \otimes C^2$ are 
distinct, the $p_i$ are positive, 
and $n \geq 2$ is minimal, i.e. $\rho$ cannot be 
represented as a mixture of fewer than $n$ pure states. 
Two parties, Alice and Bob, each have one of the spin-$1/2$ particles,
and want if possible to carry out a sequence of local quantum
operations and classical communications so as to obtain, with non-zero
probability, the singlet  
\begin{equation}
\label{singlet} 
\ket{\Psi_-} = {1 \over \sqrt{2}} ( \ket{\uparrow} \ket{\downarrow} - 
\ket{\downarrow} \ket{\uparrow} ) \, . 
\end{equation} 
As the properties of a mixed state are independent of the details of
its preparation we may, for clarity, suppose that a third party
has actually prepared one of the pure states $\ket{\psi_i}$ 
for Alice and Bob, but gives them only the statistical information 
encoded in the decomposition (\ref{m}) of $\rho$. 
In any local purification scheme, Alice and Bob carry out sequences of
POV measurements, unitary rotations, and classical communications.
They must either conclude from their measurement results that
the purification has failed, in which case they jettison the state,
or else --- if the measurements lie in some specified subset of the
possibilities --- that it has succeeded.  After any finite sequence
of operations and measurements, the density matrix takes the form 
\begin{equation}
\label{ab} 
\rho' = { {A\otimes B \rho A^\dagger\otimes B^\dagger}
\over {\Tr (A\otimes B \rho A^\dagger\otimes B^\dagger) } } \, , 
\end{equation}
where $A = A_1 ... A_{n_a}$ and $B= B_1 ... B_{n_b}$ are
the products of the positive operators and unitary
maps corresponding to Alice's and Bob's measurements and rotations. 

Now if $\rho' = \ket{\Psi_-} \bra{\Psi_-}$ then we must have that 
\begin{equation}
\label{abp}
A \otimes B \ket{\psi_i} = a_i \ket{\Psi_-} \, , 
\end{equation}
for each $i$ from $1$ to $n$ and some constants $a_i$, which need 
not all necessarily be non-zero.  
But since $A \otimes B$ maps distinct rays onto the ray $\ket{\Psi_-}$, 
it cannot have maximal rank, which means that either $A$ or $B$ 
must have rank one or zero.  This means that the $a_i$ must all be zero, 
contradicting the original hypothesis.  The same argument holds true 
for an infinite sequence of local operations, so that no 
local purification scheme involving a countable number of 
steps can produce a singlet with non-zero probability. 

Suppose now that Alice and Bob share a mixed state $\rho$ of two quantum
systems described by spaces $H_A$ and $H_B$ of dimensions 
$N_A \geq 2$ and $N_B \geq 2$.  We can write 
\begin{equation} 
\rho = \sum_{i=1}^{r} p_i \ket{\psi_i} \bra{\psi_i} \, ,
\end{equation}
where the $\ket{\psi_i}$ are orthogonal, the $p_i$ are positive, and
$r$ is the rank of $\rho$.  
Any sequence of local operations can be implemented by an operator
$A \otimes B$ as in (\ref{ab}), and if these operations 
purify $\rho$ to some state $\ket{\Psi}$ then we have 
$ A \otimes B \ket{\psi_i} = a_i \ket{\Psi}$ for each $i$ from 
$1$ to $r$ and some constants $a_i$, so that the rank 
of $ A \otimes B $ is at most $N_A N_B - r + 1$. 
If $r \geq N_A N_B - 2$, as is true for generic $\rho$, this implies that 
at least one of $A$ and $B$ has rank one or zero, so that $\ket{\Psi}$ cannot
be entangled.  Hence no local purification scheme can produce an
entangled pure state from a generic mixed state of two separated 
finite-dimensional systems.  

Since the set of possible purification operators $A \otimes B$ 
is compact, any state which cannot be completely purified to 
a maximally entangled state must have a maximal purifiability ---
according to any continuous measure --- whenever all the purification
operators produce a well-defined state.  Moreover, there must 
be some local purification for which this maximum is attained.  
For example, a generic mixed state $\rho$ of two
finite-dimensional systems has maximal rank: 
\begin{equation} 
\rho = \sum_{i=1}^{N_A N_B} p_i \ket{\psi_i} \bra{\psi_i} \, ,
\end{equation}
where the $\ket{\psi_i}$ are orthonormal.  
Thus $\Tr (  A \otimes B \rho A^\dagger \otimes B^\dagger )$ is zero
only if $A \otimes B$ is the zero operator.  Since without loss
of generality we can take $\| A \| = \| B \| = 1$, it follows that 
there is some $\epsilon ( \rho ) > 0 $ such that 
\begin{equation}
\label{maxpur}
{\rm max}_{A,B; {\ket{\psi}}}
{{ \bra{ \psi} A \otimes B \rho A^\dagger \otimes B^\dagger
    \ket{\psi}} 
  \over { \Tr (  A \otimes B \rho A^\dagger \otimes B^\dagger ) }} = 1
- \epsilon (\rho ) \, ,
\end{equation} 
where the maximum is taken over all local purification operators $A$
and $B$ and all maximally entangled pure states $\ket{\psi}$. 

It is perhaps worth stressing that this does not contradict the results of 
Bennett et al.,\cite{BBPSSW} whose scheme produces a non-zero 
fractional yield of pure Bell singlets only in the asymptotic 
limit as the number of purified pairs tends to infinity.   
Nor does the result apply to all mixed states: it is easy to 
construct examples of mixed states of lower rank which can be 
purified to singlets.  Examples of lower rank mixed states
which can be purified to arbitrary high fidelity, though not
fidelity $1$, are also known.\cite{HHH}

Finally, we consider the problem of whether the proportion of some 
entangled pure state in a mixture can be increased by a local 
purification scheme.  Suppose that 
\begin{equation}
\label{rhof} 
\rho_F = F \ket{\Psi} \bra{\Psi} + (1-F) \tilde{\rho} 
\end{equation}
is a density matrix, where $\ket{\Psi}$ is a normalised entangled state,
$0 \leq F \leq 1$, and $\tilde{\rho}$ is a density 
matrix with $\bra{\Psi} \, \tilde{\rho} \, \ket{\Psi} = 0$.  
Suppose also that there is some $F_0$ with $0 < F_0 < F$
such that $\rho_{F_0}$ is not entangled, i.e. $\rho_{F_0}$ can be 
written as a convex combination of non-entangled pure states, and
such that any state with fidelity greater than $F_0$ to $\ket{\Psi}$
is entangled. 

Then we can show that no local purification scheme can produce 
a mixed state $\rho'$ from $\rho_F$ such that the fidelity $F' = 
\bra{\Psi} \, \rho' \, \ket{\Psi}$ of $\rho'$ to 
the state $\ket{\Psi}$ is greater than $F$.  
As before, we can describe the action of any successful local purification 
scheme by (\ref{ab}).  If we consider the scheme acting on $\rho_F$, 
it produces a state of fidelity
\begin{eqnarray}
\label{ff} 
\lefteqn{F' (F) =} \qquad \\
& & {{ F | \bra{\Psi} A \otimes B \ket{\Psi} |^2 + 
           (1-F) \bra{\Psi} A \otimes B \tilde{\rho} A^\dagger \otimes
           B^\dagger \ket{\Psi} } \over 
         {\Tr ( A \otimes B ( F \ket{\Psi} \bra{\Psi} + (1-F)
           \tilde{\rho} ) A^\dagger \otimes B^\dagger ) }} \nonumber \, ,  
\end{eqnarray} 
with probability
\begin{equation} 
\Tr ( A \otimes B
 ( F \ket{\Psi} \bra{\Psi} + (1-F)
           \tilde{\rho} ) A^\dagger \otimes B^\dagger )  \, . 
\end{equation} 
Note that if this probability is non-zero for any value of 
$F$ in the range $0 < F < 1$ then it is non-zero throughout the 
range.

We have that 
\begin{eqnarray}
\label{deriv}
\lefteqn{{{d^2 } \over {d F^2}}  \delta (F)  = } \qquad \\
& & {{\beta'} \over
{ ( \Tr ( A \otimes B ( F \ket{\Psi} \bra{\Psi} + (1-F) \tilde{\rho} )
  A^\dagger \otimes B^\dagger ) )^3 } }  \nonumber \, , 
\end{eqnarray} 
where $\delta(F) = (F'(F) - F)$ and $\beta'$ is independent of
$F$, so that the sign of the second derivative is constant 
on the range $0 < F < 1$ and $\delta(F)$ is either convex, concave
or linear over the range.  Now $\delta(F) \geq -F$ and 
so approaches or exceeds zero as $F \rightarrow 0$ from above; 
similarly $\delta(F)$ approaches or is less than
zero as $F \rightarrow 1$ from below. 
By hypothesis $\delta ( F_0 ) \leq 0 $, since it is impossible to
obtain an entangled state from the unentangled state $\rho_{F_0}$. 
Hence, if there is an $F$ in the range $F_0 < F < 1$ such that $\delta(F) > 0$,
then, given the form of (\ref{ff}), $\delta (F)$ must 
have at least two extrema in the range 
$0 < F < 1$, which is inconsistent with (\ref{deriv}).  
So there can be no such $F$.  Hence no local purification scheme can
increase the fidelity of $\rho_F$ for $F \geq F_0$. 

As a special case, applying the result to a pair of spin-$1/2$
particles, taking $\ket{\Psi}$ to be the Bell singlet state 
$\ket{\Psi_-}$ and $\tilde{\rho}$ to be a linear combination 
$a \ket{\Psi_+ }\bra{\Psi_+} + b \ket{\Phi_-}\bra{\Phi_-} + 
c \ket{\Phi_+}\bra{\Phi_+}$, we obtain LMP's result that the 
fidelity of a Werner state, or any Bell-diagonal state, 
with $F>{1 \over 2}$ cannot be increased by a local purification scheme.  

In summary, the difficulty with purifying individual mixed states
is that any purification scheme must have a well-defined 
action on each of the pure states in the mixture, and hence
on all other mixtures of those states, including any non-entangled
mixtures. This imposes strong constraints, which do not arise
in purifying pure states, and which make effective purification of 
mixed states generally inconsistent with locality.  
We have used only the 
very simplest constraints here in generalising LMP's results: 
more detailed analyses would no doubt give stronger results and bounds. 

These results remove any remaining
hope that noise on a quantum channel can be completely 
countered by individual measurements on mixed states, whether or 
not the noise is rotationally symmetric.  
As LMP's results already strongly suggest,
if teleportation and similar schemes 
are to become workable, they will require either essentially
noisefree channels, or algorithms tolerant to noise, or 
technology that allows efficient collective measurements. 

\vskip15pt
\noindent{\bf Acknowledgements}
I am very grateful to Michal Horodecki, 
Noah Linden, Serge Massar, Martin Plenio and Sandu Popescu 
for helpful discussions and to the Royal Society 
for financial support.  


\end{multicols}

\end{document}